\documentstyle[amsfonts,aps]{revtex}

\begin{document}
\title{An Algebraic Formulation of Quantum Decoherence}
\author{M. A. Castagnino}
\address{Instituto de Astronom\'{i}a y F\'{i}sica del Espacio\\
Casilla de Correos 67, Sucursal 28\\
(1428) Buenos Aires, Argentina}
\author{A. R. Ord\'{o}\~{n}ez}
\address{Instituto de F\'{i}sica de Rosario\\
Av. Pellegrini 250\\
(2000) Rosario, Argentina}
\date{July 5, 2000}
\maketitle

\begin{abstract}
An algebraic formalism for quantum decoherence in systems with continuous
evolution spectrum is introduced. A certain subalgebra, dense in the
characteristic algebra of the system, is defined in such a way that
Riemann-Lebesgue theorem can be used to explain decoherence in a well
defined final pointer basis.
\end{abstract}

\section{Introduction}

Quantum decoherence was a subject of intense research in the last years \cite
{Halliwell}, \cite{Giuliani}. We have contributed to this research with
paper \cite{C}. Namely, we have presented an easy approach to explain
decoherence in a well defined {\it final pointer basis}, for quantum systems
with continuous evolution spectrum using a {\it functional} method based in
an idea of van Hove \cite{vanHowe}(\footnote{%
Some physical examples of the method can be found in \cite{MCL}.}). We have
also reobtained all the results of the method of {\it decoherence of
histories} in the final pointer basis and defined a final intrinsic
consistent set of histories \cite{Old}. But our previous approach looks like
an ``ad hoc'' one, at least in the way we presented it in papers \cite{C}
and \cite{L} (see also \cite{Anto} for a more rigorous but still ``ad hoc''
formalization).

Here we will show the general nature of this approach, introducing {\it a
rigorous mathematical formalism for our method in the framework of the
algebraic theory of quantum systems}, which is based in a {\it %
characteristic algebra }and not in a Hilbert space representation. The
importance of this ``algebraic'' (\footnote{%
It would be more precise to say ``topological- algebraic'' formalism.})
presentation lies in the fact that we may deal with systems with infinite
degrees of freedom (like an idealization of a quantum gas, or a quantum
field), where it is possible to have many non-equivalent representations of
the commutation relations, instead of just one. Therefore, the choice itself
of the adequate representation becomes a dynamical problem. Moreover, as we
generally need bounded and unbounded observables, it is better to use a {\it %
nuclear} characteristic algebra \cite{Iguri}, whose generalized
GNS-representations naturally introduce unbounded operators associated to
some riggings of the algebra with a state-dependent Hilbert space \cite
{Canadiense}. As a consequence of the {\it nuclear theorem}, our observables
will be given by nuclei or kernels -generalized matrices- and so we get a
``kernel (or nucleus) mechanics'' quite similar to the original ``matrix
mechanics''.

The physical idea behind the formalism is the following. It is well known in
the literature, that, in order to obtain decoherence, something like a
``coarse-graining'' seems to be necessary. More precisely, what produces
decoherence is a combination of certain dynamical qualities of the system
itself, together with an unavoidable restriction of the accessible
information (as it happens in a measurement, for example). So, in some way,
we have to formalize the physical fact that very frequently we do not get 
{\it the whole} relevant information about the system, but only {\it a part}
of it. Translating this into algebraic terms, we could say that sometimes it
is not possible to use neither {\it all }the characteristic algebra ${\cal A}
$ of our system, nor {\it all }the corresponding symmetric elements or
observables, but just a certain {\it subalgebra\ }and its subset of
observables. Of course, in principle there are infinite ways of making that
choice of a subalgebra. But the idea is to do it in such a way that it would
give us the possibility of the annihilation -at least for $t\rightarrow
+\infty $- of all the ``off-diagonal terms'' of the states by an application
of the Riemann-Lebesgue theorem.

In our subalgebra, that will be called {\it the van Hove algebra} ${\cal A}%
_{vH}$, we will have two kinds of observables:

i.- observables $a$ measuring the ``diagonal terms'', that commute with the
Hamiltonian and can be given by a ``singular'' or ``semiregular'' kernel of
the form: $a(\omega )\delta (\omega -\omega ^{\prime })$ (\footnote{%
Where $\omega $ is the energy, i. e. an eigenvalue of the hamiltonian
generating the evolution.})

ii.- observables whose mean value goes to zero when $t\rightarrow +\infty $
represented by ``regular'' kernels of the type: $a_{r}(\omega ,\omega
^{\prime })$

(Here $a(\omega )$ and $a_{r}(\omega ,\omega ^{\prime })$ are not
distributions, but ``ordinary'' functions -respectively of one and two
energy variables $\omega ,\omega ^{\prime }$-, whose mathematical properties
are sufficient in order to use the Riemann-Lebesgue theorem, in its two
variables case, in such a way that the off-diagonal terms $a_{r}(\omega
,\omega ^{\prime })$ vanish when $t\rightarrow +\infty $).

As any singular kernel can be aproximated by a regular one, the van Hove
algebra ${\cal A}_{vH}$ will be dense in ${\cal A}$. Therefore, ${\cal A}%
_{vH}$ will not be {\it mathematically} complete, but it will be {\it %
physically} ``complete'', in the sense of having all the required physical
properties up to any order, and therefore being experimentally
indistinguishable from ${\cal A}$.

The paper is organized as follows:

In section 2 we review some basic and remarkable facts about the nuclear
*-formalism for quantum mechanics.

In section 3 we introduce the nuclear *-algebra ${\cal L}\left( {\cal S(}%
{\Bbb R}^{+}{\Bbb )}\right) $, associated with the Schwartz space ${\cal S(}%
{\Bbb R}^{+}{\Bbb )}$.

In section 4 we first introduce a simple but very illustrative example: a
quantum system whose characteristic algebra is ${\cal L}\left( {\cal S(}%
{\Bbb R}^{+}{\Bbb )}\right) $, and where the CSCO (complete set of commuting
observables) is just the hamiltonian $H$. We define its van Hove algebra,
and show how the evolution of the quantum system yields decoherence on it.

In section 5, we consider a more general quantum system, with a CSCO given
by $\left\{ H,O_{1},O_{2},...,O_{N}\right\} $ whose spectrum is $\Lambda
\subset {\Bbb R}^{N+1}$. We show how to obtain its characteristic nuclear
*-algebra ${\cal L}\left( {\cal S}_{\Lambda }\left( {\Bbb R}^{N+1}\right)
\right) $ and we generalize the van Hove algebra, yielding decoherence.

Finally, in section 6 we draw our main conclusions.

\section{Quantum mechanics in a nuclear *-algebra ${\cal A}$}

\subsection{Dynamics of ${\cal A}$}

Let us consider a nuclear *-algebra ${\cal A}$ (\cite{Iguri} \cite{Borchers} 
\cite{Treves} \cite{Pietsch}) as the {\bf characteristic algebra} of a
quantum system. This amounts to say that we can describe any physical
property of the system in terms of mathematical objects related to ${\cal A}$%
. For example, all the commutation relations of the observables (\footnote{%
Bounded or unbounded, instead of only the bounded one's as happen within the
B*-algebras.}) of the system can be expressed or represented in terms of the
commutators of the algebra: $[a,b]:=ab-ba$, all the physical symmetries can
be represented as inner automorphism groups of the algebra, etc (\footnote{%
We must remark that the correspondence between systems and characteristic
algebras is not generally injective: there could be unequivalent physical
systems with the same algebra. (This is not a surprise, because the same
happens with the Hilbert space formalism).}). In particular, there exists a
one parameter group of unitary inner automorphism 
\begin{equation}
{\cal U}_{t}:{\cal A}\rightarrow {\cal A}\;/\;{\cal U}%
_{t}(a)=u_{t}au_{t}^{-1}  \label{1.1}
\end{equation}
with $u_{t}$ {\it unitary} (i. e. : $u_{t}^{*}=u_{t}^{-1}$), representing
the temporal evolution of the system. Usually $u_{t}=e^{-iHt}$ where $H$ is
the hamiltonian operator, {\it which will be supposed to have an absolutely
continuous spectrum in a real interval contained in }${\Bbb [}0,{\Bbb %
+\infty )}$. This hypotesis is crucial in order to use the Riemann-Lebesgue
theorem. The spectrum could also contain an eigenvalue, corresponding to a
bound state (for example, a ground state \cite{C}).

Remark: here the concept of spectrum can be either the usual one -if we are
dealing with a finite degree of freedom theory with a fixed Hilbert space-,
or its generalization to nuclear *-algebras. In fact, it will be an
essential hypothesis for any nuclear *-algebra in order to be physically
admisible as the characteristic algebra of a quantum system, to have the
``right'' spectral properties. Namely, if it is the algebra of a system
having a well known Hilbert space representation, its symmetric elements
must have real spectrum identical to its corresponding Hilbert space
representatives, etc. In particular, it can be a {\bf nuclear b*-algebra},
that is to say a proyective limit of B*-algebras, as in \cite{Iguri}, but it
could be more general.

${\cal U}_{t}$ preserves the * operation: if $a^{*}=a$ then 
\begin{equation}
\left( u_{t}au_{t}^{-1}\right) ^{*}=u_{t}au_{t}^{-1}  \label{1.2}
\end{equation}

In other words, the * operation is also an automorphism in ${\cal A}_{S}$ ,
the real space of {\it symmetric} operators of ${\cal A}$, that is to say,
of all $a\in {\cal A}$ such that $a^{*}=a$.

\subsection{Observables in ${\cal A}$}

${\cal A}_{S}$ is closed in ${\cal A}$, and therefore it is a real nuclear
space that will be called the {\bf space of observables }of the system. If $%
{\cal A}$ is not commutative, the product of two symmetric elements will not
be symmetric, and therefore ${\cal A}_{S}$ will not be a subalgebra of $%
{\cal A}$.

Let us consider a CSCO of the system. Then, there exists a unique minimal
closed abelian subalgebra with unity containing it, called {\bf the abelian
subalgebra generated }by this CSCO, that will be labelled $\widehat{{\cal A}}
$. Obviously, we have the relations:

\begin{equation}
\widehat{{\cal A}}\subset {\cal A}_{S}\subset {\cal A}  \label{1.3}
\end{equation}
$\widehat{{\cal A}}$ is a complete subspace and a subalgebra of ${\cal A}$.
As we are focused in decoherence, that is not a ``covariant'' subject (%
\footnote{%
In the sense that we must have a privileged observable, namely the
hamiltonian $H$. Also the final pointer basis will depend on $H$ and the
initial conditions (see \cite{C})}), all the CSCO that will be considered
here contain the hamiltonian $H.$

We can also define the {\bf cone of positive observables} $a\in {\cal A}_{S}$
as: 
\begin{eqnarray}
{\cal A}_{S+} &=&\{a\in {\cal A\;}/\;\exists \text{ }b\in {\cal A}:\text{ }%
a=b^{*}b\}  \nonumber \\
&=&\{a\in {\cal A\;}/\;\exists \text{ }b\in {\cal A}_{S}:\text{ }a=b^{2}\}
\label{1.4}
\end{eqnarray}

This set is a cone since any linear combination of its elements with
positive coefficients belong to ${\cal A}_{S+}.$

\subsection{Convex of states}

The space of states is: 
\begin{equation}
N\left( {\cal A}_{S+}^{\prime }\right) =\{\rho \;/\;\rho \in {\cal A}%
_{S}^{\prime }\wedge \rho \geq 0\wedge \rho ({\Bbb I)=}1\}  \label{1.5}
\end{equation}
Precisely: $\rho \geq 0$ means $\rho (a)\geq 0$ for all $a\in {\cal A}_{S+}$
, or equivalently $\rho (b^{*}b)\geq 0$ for all $b\in {\cal A}.$ As in \cite
{C}, the generalization of the trace is $Tr(\rho ):=\rho ({\Bbb I}) $ where $%
{\Bbb I}$ is the identity operator of algebra ${\cal A}$. So $\rho ({\Bbb I)=%
}1$ is the normalization (or unit trace) condition. Clearly, $N\left( {\cal A%
}_{S+}^{\prime }\right) $ is a convex subset of ${\cal A}_{S}^{\prime }$ and
therefore inherited its topology. The particular states that are {\it %
extremals} of this convex constitute the subset of {\bf pure states}. The
finite convex combinations of pure states are called {\bf mixed states}. The
remaining states (functionals that cannot be represented in these two ways,
but are expresables as {\it integrals }or ``infinite{\it \ }conbinations''
of pures states, by a generalization of the Krein-Milman theorem \cite{Iguri}
\cite{Commun} \cite{Naimark} ) are called {\bf generalized states }\cite{L}%
{\bf .}

There is also a trace preserving group of automorphism in the states space $%
N\left( {\cal A}_{S+}^{\prime }\right) $. In fact, the evolution
automorphism over the observables of eq. (\ref{1.1}), induces the following
group in the dual space of the algebra: ${\cal U}_{t}^{\prime }:{\cal A}%
^{\prime }\rightarrow {\cal A}^{\prime }$ defined as:

\begin{equation}
\left( {\cal U}_{t}^{\prime }\rho \right) (a):=\rho [{\cal U}_{t}(a)]=\rho
(u_{t}au_{t}^{-1})  \label{1.6}
\end{equation}
in another notation 
\begin{equation}
\left( {\cal U}_{t}^{\prime }\rho \right) (a)=\rho _{t}(a)=\rho [{\cal U}%
_{t}(a)]=\rho (a_{t})  \label{1.7}
\end{equation}

This evolution preserve the trace: $\rho _{t}({\Bbb I)=}1$ and the energy $%
\rho _{t}(H{\Bbb )=}$\ constant. Therefore it corresponds to the {\it %
Schr\"{o}dinger picture} while (\ref{1.1}) corresponds to the {\it %
Heisenberg picture}.

Since $\widehat{{\cal A}}$ is a subalgebra of ${\cal A}$ we can consider the
states that correspond to $\widehat{{\cal A}\text{,}}$ namely the set of
positive normalized functionals $N\left( \widehat{{\cal A}}_{+}^{\prime
}\right) $ that will be called the convex of {\bf reduced states }with
respect to $\widehat{{\cal A}}$. By a classical theorem of M. G. Krein (\cite
{Naimark}, pag. 63, Theorem 2), and the fact that the identity is an
internal element of any CSCO, any reduced state can be extended to the whole 
$N\left( {\cal A}_{S+}^{\prime }\right) .$

Since $\widehat{{\cal A}}$ is commutative, its elements can be considered
just as ${\cal A}$-valued functions of the observables of the corresponding
CSCO, and its states -belonging to $N\left( \widehat{{\cal A}}_{+}^{\prime
}\right) $- as ${\Bbb C}$-valued functions of them. Therefore in some way
they are ``quasiclassical''. In fact, the final state of a quantum
measurement of the observables in the CSCO will be a corresponding reduced
state (all the information content of the state concerning other
non-commuting observables being eventually lost). So, we can describe the
process of decoherence as a kind of temporal ``homotopy'' $h_{t}$ mapping of 
$N\left( {\cal A}_{S+}^{\prime }\right) $ into $N\left( \widehat{{\cal A}}%
_{+}^{\prime }\right) $ as $t\rightarrow +\infty $. At this moment we want
to remark that the set of pure states of the system is defined as a
particular component of the structure of the algebra, and therefore it
cannot be altered by any unitary inner automorphism, including the temporal
evolution of the system. In other words, the temporal ``homotopy'' $h_{t}$
must preserve ``the boundary'' -the pure states- of the convex of states.
So, the evolution cannot possible act as a process of decoherence {\it on
the whole} convex of states. Nevertheless, as we said in the introduction,
this can happen{\it \ in a subset} of the mixed or generalized states:
precisely in the {\bf Van Hove (or decoherent) states,} that will be defined
in the following sections.

\section{The nuclear *-algebra ${\cal L}\left( {\cal S}\left( {\Bbb R}%
^{+}\right) \right) $}

\subsection{The nuclear algebra ${\cal L}\left( {\cal S}\left( {\Bbb R}%
^{+}\right) \right) $}

Let ${\cal S(}{\Bbb R}^{+}{\Bbb )}$ be the closed subspace of the Schwarz
space ${\cal S(}{\Bbb R)}$ \cite{Rudin} consisting of all $C^{\infty }{\Bbb [%
}0,{\Bbb +\infty )}$ functions $f$ such that: 
\begin{equation}
p_{n,m}(f)=\sup_{x\in {\Bbb [}0,+\infty )}(1+x^{2})^{n}|D^{m}f(x)|<\infty
\label{2.1}
\end{equation}
i.e. the functions $f$ and all its derivatives go to zero when $x\rightarrow
+\infty $ faster than the inverse of any polynomial function. The topology
is defined by the $p_{n,m}(f),$ in the sense that $f_{i}\rightarrow f$ if $%
\forall n,m$ :$\;p_{n,m}(f_{i}\rightarrow f)\rightarrow 0\;$as $i\rightarrow
+\infty $. This is a closed subspace of ${\cal S(}{\Bbb R)}$ and hence a
complete metrizable nuclear space in itself. In the space of distribution or
functionals ${\cal S}^{\prime }{\cal (}{\Bbb R}^{+}{\Bbb )}$ we consider the
strong topology, that is to say, the locally convex topology obtained from
the systems of seminorms 
\begin{equation}
p_{B}^{\prime }(\alpha )=\sup \{\left| \alpha (g)\right| \;/\;g\in B\}
\label{2.2}
\end{equation}
for any fundamental system of bounded subsets ($B)$ of ${\cal S(}{\Bbb R}^{+}%
{\Bbb )}$. Then ${\cal S}^{\prime }{\cal (}{\Bbb R}^{+}{\Bbb )}$ is a
nuclear space because ${\cal S(}{\Bbb R}^{+}{\Bbb )}$ is nuclear and
metrizable \cite{Treves} \cite{Pietsch}.

Analogously, we can define ${\cal S(}{\Bbb R}^{+}\times {\Bbb R}^{+}{\Bbb )}$
and its strong dual ${\cal S}^{\prime }{\cal (}{\Bbb R}^{+}\times {\Bbb R}%
^{+}{\Bbb )}$, constituted by {\it kernels distributions,} i. e.,
distributions in two (here non-negative) variables.

Now, it is known (see formula (51.7) of \cite{Treves}) that 
\begin{equation}
{\Bbb {\cal S}}^{\prime }{\Bbb {\cal (}R^{+})}\widehat{{\Bbb \otimes }}{\Bbb 
{\cal S}}^{\prime }{\Bbb {\cal (}R^{+})\cong }{\cal S}^{\prime }{\cal (}%
{\Bbb R}^{+}\times {\Bbb R}^{+}{\Bbb )\cong }{\cal L}\left( {\cal S(}{\Bbb R}%
^{+}{\Bbb )\,},{\cal S}^{\prime }{\cal (}{\Bbb R}^{+}{\Bbb )}\right)
\label{2.3}
\end{equation}
where $\widehat{{\Bbb \otimes }}$ denotes the completion of the tensor
product (carrying its projective $\pi $-topology, or its equicontinuous $%
\varepsilon $-topology, for here they are equivalent). The last of these two
isomorphisms is defined as follows. If $K\in {\cal S}^{\prime }{\cal (}{\Bbb %
R}^{+}\times {\Bbb R}^{+}{\Bbb )}$, to any $f\in {\cal S(}{\Bbb R}^{+}{\Bbb )%
}$, we can associate $\alpha \in {\cal S}^{\prime }{\cal (}{\Bbb R}^{+}{\Bbb %
)}$ such that 
\begin{equation}
\forall g\in {\cal S(}{\Bbb R}^{+}{\Bbb )}:\alpha (g)=K\left( g\otimes
f\right)  \label{2.4}
\end{equation}

It is traditional, specially in ``physicists' jargon'' , to write this as: 
\begin{equation}
K(x,x^{\prime })\leftrightarrow \alpha :\left[ \alpha (f)\right] (x)=\int
K(x,x^{\prime })f(x^{\prime })dx^{\prime }  \label{2.5}
\end{equation}

Then, the isomorphism is given by the linear map 
\begin{equation}
K\leftrightarrow a:a(f)=\alpha  \label{2.6}
\end{equation}

Taking into account the relations 
\begin{equation}
{\cal S(}{\Bbb R}^{+}{\Bbb )\hookrightarrow }L^{2}{\cal (}{\Bbb R}^{+}{\Bbb %
)\hookrightarrow }{\cal S}^{\prime }{\cal (}{\Bbb R}^{+}{\Bbb )}  \label{2.7}
\end{equation}
defined by the injections 
\begin{equation}
f\hookrightarrow [f]\hookrightarrow \alpha _{f}(g):=\int f(x)g(x)dx
\label{2.8}
\end{equation}
where $[f]$ is the class of functions that are a. e. (``almost everywhere'')
equal to $f$, and thinking similarly to reference \cite{Treves} pages
532-533, we will say that:

i) the kernel $K$ or its associated map $a$ are {\it semiregular in }$x$ if $%
a$ maps ${\cal S(}{\Bbb R}_{x}^{+}{\Bbb )}$ into ${\cal S(}{\Bbb R}%
_{x^{\prime }}^{+}{\Bbb )}$, and not only into ${\cal S}^{\prime }{\cal (}%
{\Bbb R}_{x^{\prime }}^{+}{\Bbb )}$.

ii) the kernel $K$ or its associated map $a$ are {\it semiregular in }$%
x^{\prime }${\it \ }if the transpose $a^{\prime }$ of $a$ maps ${\cal S(}%
{\Bbb R}_{x}^{+}{\Bbb )}$ into ${\cal S(}{\Bbb R}_{x^{\prime }}^{+}{\Bbb )} $%
, and not only into ${\cal S}^{\prime }{\cal (}{\Bbb R}_{x^{\prime }}^{+}%
{\Bbb )}$.

iii) $K$ is a {\it regular} {\it kernel} if it is the regular distribution
given by a function $K(x,x^{\prime })$ of ${\cal S(}{\Bbb R}_{x}^{+}\times 
{\Bbb R}_{x^{\prime }}^{+}{\Bbb )}$.

For example, the Dirac's delta $\delta (x-x^{\prime })$ is semiregular in
both $x$ and $x^{\prime }$, because it is symmetric in $x$ and $x^{\prime }$%
, but obviously is not a regular kernel. So, ${\cal S}^{\prime }\left( {\Bbb %
R}^{+}{\Bbb \times R}^{+}\right) $ has a lot of physically important
kernels. In fact it has {\it too many}, to the extent of not being an
algebra because of the well known product problem of the distributions. In
order to avoid this obstacle, we restrict ourselves to 
\begin{equation}
{\cal S}^{\prime }{\cal (}{\Bbb R}^{+}{\Bbb )}\widehat{{\Bbb \otimes }}{\cal %
S(}{\Bbb R}^{+}{\Bbb )\subset \;}{\cal S}^{\prime }{\cal (}{\Bbb R}^{+}{\Bbb %
)}\widehat{{\Bbb \otimes }}{\cal S}^{\prime }{\cal (}{\Bbb R}^{+}{\Bbb %
)\cong \;}{\cal S}^{\prime }{\cal (}{\Bbb R}^{+}\times {\Bbb R}^{+}{\Bbb )}
\label{2.9}
\end{equation}

This amounts the restriction to the algebra: 
\begin{equation}
{\cal A}={\cal L}\left( {\cal S(}{\Bbb R}^{+}{\Bbb )}\right) ={\cal L}\left( 
{\cal S(}{\Bbb R}^{+}{\Bbb )\,},{\cal S(}{\Bbb R}^{+}{\Bbb )}\right) \subset 
{\cal L}\left( {\cal S(}{\Bbb R}^{+}{\Bbb )\,},{\cal S}^{\prime }{\cal (}%
{\Bbb R}^{+}{\Bbb )}\right)  \label{2.10}
\end{equation}
where the continuity is defined in the sense of ${\cal S(}{\Bbb R}^{+}{\Bbb )%
}$. In fact, 
\begin{equation}
{\cal L}\left( {\cal S(}{\Bbb R}^{+}{\Bbb )}\right) \cong {\cal S}^{\prime }%
{\cal (}{\Bbb R}^{+}{\Bbb )}\widehat{{\Bbb \otimes }}{\cal S(}{\Bbb R}^{+}%
{\Bbb )\subset }{\cal S}^{\prime }{\cal (}{\Bbb R}^{+}{\Bbb )}\widehat{{\Bbb %
\otimes }}{\cal S}^{\prime }{\cal (}{\Bbb R}^{+}{\Bbb )}  \label{2.11}
\end{equation}

In particular, ${\cal L}\left( {\cal S(}{\Bbb R}^{+}{\Bbb )}\right) $ is a
nuclear algebra. Thus, any element of ${\cal L}\left( {\cal S(}{\Bbb R}^{+}%
{\Bbb )}\right) $ can be considered as a ``generalized matrix'' with a lower
index corresponding to ${\cal S}^{\prime }{\cal (}{\Bbb R}^{+}{\Bbb )}$ and
an upper one corresponding to ${\cal S(}{\Bbb R}^{+}{\Bbb )}$.

So, we realize the clearest and most intuitive idea of nuclearity, based in
the {\bf nuclear theorem, }and{\bf \ }the very etymology of ``nuclear
algebras'': they are algebras of nuclei or kernels that are multiplied as
generalized matrices \cite{Treves} \cite{Pietsch}. In fact, let us define a
linear and continuous mapping: 
\begin{equation}
B:{\cal S}^{\prime }{\cal (}{\Bbb R}^{+}{\Bbb )\times }{\cal S(}{\Bbb R}^{+}%
{\Bbb )\rightarrow C}  \label{2.12}
\end{equation}

As any continuous function from a nuclear space into a Banach space is
nuclear, this mapping is nuclear, and according to the nuclear theorem,
there exists a linear and continuous mapping: 
\begin{equation}
K:{\cal S}^{\prime }{\cal (}{\Bbb R}^{+}{\Bbb )\otimes }{\cal S(}{\Bbb R}^{+}%
{\Bbb )\rightarrow C}  \label{2.13}
\end{equation}
such that 
\begin{equation}
B(\alpha ,g)=K(\alpha \otimes g),\qquad \forall \alpha \in {\cal S}^{\prime }%
{\cal (}{\Bbb R}^{+}{\Bbb )}\text{, }\forall g\in {\cal S(}{\Bbb R}^{+}{\Bbb %
)}  \label{2.14}
\end{equation}

Now let $a\in {\cal A=}{\cal L}\left( {\cal S(}{\Bbb R}^{+}{\Bbb )}\right) $%
. For any $\alpha \in {\cal S}^{\prime }{\cal (}{\Bbb R}^{+}{\Bbb )}$ and
any $g\in {\cal S(}{\Bbb R}^{+}{\Bbb )}$, we can define a bilinear and
continuous mapping $B_{a}$: 
\begin{equation}
B_{a}(\alpha ,g):=\alpha [a(g)]  \label{2.15}
\end{equation}
and since ${\cal A=L}\left( {\cal S(}{\Bbb R}^{+}{\Bbb )}\right) $ is a
nuclear algebra $\exists $ $K_{a}$ such that: 
\begin{equation}
K_{a}(\alpha \otimes g)=\alpha [a(g)]  \label{2.16}
\end{equation}

In the ``physicists' jargon'' it is: 
\begin{equation}
K_{a}(\alpha \otimes g)=\int \int \alpha (x)K_{a}(x,x^{\prime })g(x^{\prime
})dxdx^{\prime }  \label{2.17}
\end{equation}
{\bf Examples}. It is easy to verify that:

1.-If $a={\Bbb I}$ then $K_{a}(\alpha \otimes g)=\alpha (g)$ (in finite
dimension it would be the contraction, whose matrix is the Kronecker delta),
and $K_{a}(x,x^{\prime })=\delta (x-x^{\prime }).$

2.-If $[a(g)](x):=xg(x),$ then $K_{a}(x,x^{\prime })=x\delta (x-x^{\prime })$
is simiregular in $x$ and $x^{\prime }.$ More generally, when $%
[a(g)](x):=f(x)g(x),$ then $K_{a}(x,x^{\prime })=f(x)\delta (x-x^{\prime })$
is simiregular in $x$ and $x^{\prime }$

3.-If $K(x,x^{\prime })$ is a regular kernel, for any $g(x)\in {\cal S(}%
{\Bbb R}^{+}{\Bbb )}$ the function 
\begin{equation}
\text{ }f(x):=\int K(x,x^{\prime })g(x^{\prime })dx^{\prime }  \label{2.18}
\end{equation}
belongs to ${\cal S(}{\Bbb R}^{+}{\Bbb )}$, and therefore we can define an
operator $a\in {\cal L}\left( {\cal S(}{\Bbb R}^{+}{\Bbb )}\right) $, by: 
\begin{equation}
\left[ a(g)\right] (x):=f(x)  \label{2.19}
\end{equation}

4.-If $K(x,x^{\prime })$ is a general distribution$,$ then eq. (\ref{2.18})
defines a tempered distribution (because in that case $f(x)$ is not
necessarily a Schwarz function, moreover, it may not even be a function).
Thus, we return to the correspondence 
\begin{equation}
K(x,x^{\prime })\mapsto \alpha _{x}:\left[ \alpha _{x}\left( g(x^{\prime
})\right) \right] (x):=\int K(x,x^{\prime })g(x^{\prime })dx^{\prime }
\label{2.20}
\end{equation}
which (as was already shown), is a non-surjective injection 
\begin{equation}
\left( {\cal S}^{\prime }{\cal (}{\Bbb R}^{+}{\Bbb )\otimes }{\cal S(}{\Bbb R%
}^{+}{\Bbb )}\right) ^{\prime }\rightarrow {\cal L}\left( {\cal S(}{\Bbb R}%
^{+}{\Bbb )\,},\,{\cal S}^{\prime }{\cal (}{\Bbb R}^{+}{\Bbb )}\right)
\label{2.21}
\end{equation}

\subsection{The star operation in ${\cal L}\left( {\cal S(}{\Bbb R}^{+}{\Bbb %
)}\right) $}

We will show that ${\cal L}\left( {\cal S(}{\Bbb R}^{+}{\Bbb )}\right) $ is
also a *-algebra. We know that $\forall a\in {\cal L}\left( {\cal S(}{\Bbb R}%
^{+}{\Bbb )}\right) $ there exists the {\it dual or transpose map} $%
a^{\prime }:{\cal S}^{\prime }{\cal (}{\Bbb R}^{+}{\Bbb )\rightarrow }{\cal S%
}^{\prime }{\cal (}{\Bbb R}^{+}{\Bbb )}$ such that: 
\begin{equation}
\left[ a^{\prime }(\alpha )\right] (g):=\alpha \left[ a(g)\right]
\label{3.3}
\end{equation}

Similarly we can define $a^{\dagger }:{\cal S}^{\prime }{\cal (}{\Bbb R}^{+}%
{\Bbb )\rightarrow }{\cal S}^{\prime }{\cal (}{\Bbb R}^{+}{\Bbb )}$ such
that: 
\begin{equation}
\left[ a^{\dagger }(\alpha )\right] (g):=\alpha \left[ \overline{a\left( 
\overline{g}\right) }\right]  \label{3.4}
\end{equation}

Let us define: 
\begin{equation}
a^{*}=a^{\dagger }|_{{\cal S(}{\Bbb R}^{+}{\Bbb )}}  \label{3.5}
\end{equation}
with the restriction according to eq. (\ref{2.8}). Then 
\begin{equation}
\left[ \overline{a(\overline{g})}\right] (x)=\int \overline{%
K_{a}(x,x^{\prime })}g(x^{\prime })dx^{\prime }  \label{3.6}
\end{equation}
and 
\begin{equation}
\lbrack a^{*}(f)](x)=\left[ a^{\dagger }(\alpha _{f})\right] (x)=\int
f(x^{\prime })K_{a^{*}}(x^{\prime },x)dx^{\prime }  \label{3.7-1}
\end{equation}

But according to eq. (\ref{3.4}), we have 
\begin{eqnarray}
\left[ a^{\dagger }(\alpha _{f})\right] (g) &=&\int \left[ a^{\dagger
}(\alpha _{f})\right] (x)g(x)dx  \nonumber \\
&=&\int \int f(x^{\prime })K_{a^{*}}(x^{\prime },x)g(x)dx^{\prime }dx 
\nonumber \\
&=&\int \left[ \overline{a(\overline{f})}\right] (x)g(x)dx  \nonumber \\
&=&\int \int f(x^{\prime })\overline{K_{a}(x,x^{\prime })}g(x)dx^{\prime }dx
\label{3.7-2}
\end{eqnarray}

Thus 
\begin{equation}
K_{a^{*}}(x^{\prime },x)=\overline{K_{a}(x,x^{\prime })}  \label{3.8}
\end{equation}
and therefore the star operation is the conjugation followed by the
transposition as in the case for ordinary matrices.

By its definition, it is clear that this * operation is a continuous
antihomomorphism of algebras. And as ${\cal S(}{\Bbb R}^{+}{\Bbb )}$ is a
reflexive space (because it is a Montel space \cite{Treves}, p. 376), it is
involutive.

Summarizing, ${\cal L\left( S({\Bbb R}^{+}{\Bbb )}\right) }$ is a complete
nuclear *-algebra.

\section{Quantum mechanics in ${\cal A=L\left( S({\Bbb R}^{+}{\Bbb )}\right) 
}$}

\subsection{The simplest example}

For the sake of simplicity, let us consider a physical system whose
Hamiltonian has $\left[ 0,+\infty \right) $ as absolutely continuous
spectrum, and such that $\{H\}$ is a CSCO generating $\widehat{{\cal A}}$
(we will generalize this CSCO in the next section)$.$ Clearly, 
\[
\text{Clausure in }{\cal A}\text{ of }\left\{ \widehat{a}\in {\cal A\;}/\;K_{%
\widehat{a}}(\omega ,\omega ^{\prime })=\widehat{a}(\omega )\delta (\omega
-\omega ^{\prime })\wedge \widehat{a}(\omega )\in {\Bbb C}[\omega ]\right\}
\subset \widehat{{\cal A}} 
\]
(${\Bbb C}[\omega ]=$ set of all polinomial functions in $\omega $ with
complex coefficients). As the left hand side of this inclusion relation {\it %
already is} a closed commutative subalgebra with unit of ${\cal A} $
containing $\{H\}$, the inclusion of above must be an equality.

Then, ${\cal A=L\left( S({\Bbb R}^{+}{\Bbb )}\right) }$ is its natural
characteristic algebra, and according to the previous section (example 2.-),
the particular semiregular kernels of type $\widehat{a}(\omega )\delta
(\omega -\omega ^{\prime })$ correspond to the elements of $\widehat{{\cal A}%
}$.

\subsection{The van Hove algebra}

Let us consider the quotient ${\cal A\;}$/$\;\widehat{{\cal A}}$ which is a
vector space (but not a subalgebra since $\widehat{{\cal A}}$ is not an
ideal). Let us call 
\begin{equation}
{\cal A\;}/\;\widehat{{\cal A}}:={\cal V}_{nd}  \label{5.1}
\end{equation}
the ``non-diagonal'' vector space. Then, if $[a]\in {\cal V}_{nd}$ and $a\in 
{\cal A}$: 
\begin{equation}
\lbrack a]=\widehat{{\cal A}}+a  \label{5.2}
\end{equation}
so 
\begin{equation}
{\cal A=}\widehat{{\cal A}}+{\cal V}_{nd}  \label{5.3}
\end{equation}
where the last ``+'' symbol is not a direct sum, since we can add and
substract an arbitrary $a\in \widehat{{\cal A}}$ to each term of the r. h.
s. But we can turn ``+'' into a ``$\oplus "$ if we restrict ourselves to a
smaller (but dense, and so physically equivalent) subalgebra of ${\cal A}$.

In general, the kernels of ${\cal V}_{nd}$ are tempered distributions. Now,
let us restrict these last kernels to be just regular ones, constituting a
space ${\cal V}_{r}\subset {\cal V}_{nd}.$ Then we can define the {\bf van
Hove algebra} as: 
\begin{eqnarray}
{\cal A}_{vH} &=&\widehat{{\cal A}}\oplus {\cal V}_{r}  \nonumber \\
&=&\{a\in {\cal A\;}/\;K_{a}(\omega ,\omega ^{\prime })=\widehat{a}(\omega
)\delta (\omega -\omega ^{\prime })+a_{r}(\omega ,\omega ^{\prime })\}
\label{5.4}
\end{eqnarray}
where $\widehat{a}(\omega ),$ and $a_{r}(\omega ,\omega ^{\prime })$ are
``regular'' functions, in the sense of being endowed with the properties
listed in the Appendix of paper \cite{L} (they were chosen to be {\it natural%
} from a physical point of view, and {\it sufficient} in order to satisfy
the mathematical hypotesis of the Riemann-Lebesgue theorem, {\it in the two
variables case}), namely:

1.- $\widehat{a}(\omega )\in {\cal S(}{\Bbb R}^{+}{\Bbb )}$

2.- $a_{r}(\omega ,\omega ^{\prime })\in {\cal S}\left( {\Bbb R}^{+}\times 
{\Bbb R}^{+}\right) $

Now we have a $\oplus $ because a kernel cannot be a Dirac's $\delta $ and a
regular function at the same time. It is easy to prove that ${\cal A}_{vH} $
is a *-subalgebra of ${\cal A},$ and hence a nuclear algebra in itself, but
a non complete one (because it is not closed). Nevertheless it is dense in $%
{\cal A}$ (because as it is well known, any distribution is a limit of
regular functions). Anyhow, here the non-completeness is not a problem,
because we are not interested in taking general limits in ${\cal A}_{vH}$.
Let us denote $\widehat{a}$, $a_{r}\in {\cal A}$ the linear operarors whose
kernels are $\widehat{a}(\omega )\delta (\omega -\omega ^{\prime })$ and $%
a_{r}(\omega ,\omega ^{\prime })$, respectively. Then we can write 
\[
a=\widehat{a}+a_{r} 
\]

Now consider the time evolution within the van Hove algebra. If $a\in {\cal A%
}_{vH}$ then 
\begin{equation}
{\cal U}_{t}(a)={\cal U}_{t}(\widehat{a}+a_{r})=\widehat{a}+{\cal U}%
_{t}(a_{r})  \label{5.5}
\end{equation}
where $\widehat{a}\in \widehat{{\cal A}}$ and $a_{r}\in {\cal V}_{r}.$ The
last equation shows the most important characteristic of the semiregular and
regular parts: the semiregular part $\widehat{a}$ is {\bf invariant} under
time evolution while the regular $a_{r}$ is{\bf \ fluctuating}.

As we are particulary interested in observables, i. e. symmetric elements of
the algebra, we define the {\it space of van Hove observables}, 
\begin{equation}
{\cal A}_{vHS}:=\left\{ a\in {\cal A}_{vH}{\cal \;}/\;a^{*}=a\right\}
\label{5.5.1}
\end{equation}

In particular, $a\in {\cal A}_{vHS}$ implies that:

3.- $\widehat{a}(\omega )$ is a {\it real}-valued regular function of ${\cal %
S(}{\Bbb R}^{+}{\Bbb )}$, and

4.- the regular term is {\it hermitian}, i. e., it verifies

\[
a_{r}(\omega ,\omega ^{\prime })=\overline{a_{r}(\omega ^{\prime },\omega )} 
\]

\subsection{The van Hove states}

Now, we are going to define the states. First, let 
\begin{equation}
{\cal A}_{vH}^{\prime }=\widehat{{\cal A}}^{\prime }\oplus {\cal V}%
_{r}^{\prime }  \label{5.6.1}
\end{equation}
where $\widehat{{\cal A}}^{\prime }$ is the topological dual of $\widehat{%
{\cal A}}$, but ${\cal V}_{r}^{\prime }$ is just a notation for the set of
all functionals $\rho _{r}\in {\cal A}^{\prime }$, whose kernels $\rho
_{r}(\omega ,\omega ^{\prime })$ satisfy:

1'.- $\rho _{r}(\omega ,\omega ^{\prime })a_{r}(\omega ^{\prime },\omega
)\in {\cal S}\left( {\Bbb R}^{+}\times {\Bbb R}^{+}\right) $ for any $%
a_{r}\in {\cal V}_{r}.$

Then, if $a\in {\cal A}_{vH}$ we have: 
\begin{equation}
\rho (a)=\int\limits_{0}^{+\infty }\widehat{\rho }(\omega {\Bbb )}\widehat{a}%
(\omega )d{\Bbb \omega }+\int\limits_{0}^{+\infty }\int\limits_{0}^{+\infty
}\rho _{r}(\omega ,\omega ^{\prime })a_{r}(\omega ^{\prime },\omega )d\omega
^{\prime }d{\Bbb \omega }  \label{5.9}
\end{equation}

If they also satisfy:

2'.- $\widehat{\rho }(\omega )$ is a {\it real}-valued regular function of $%
{\cal S(}{\Bbb R}^{+}{\Bbb )}$, and

3'.- the regular term is {\it hermitian}, i. e., it verifies 
\[
\rho _{r}(\omega ,\omega ^{\prime })=\overline{\rho _{r}(\omega ^{\prime
},\omega )} 
\]

\noindent they belong to

\begin{equation}
{\cal A}_{vHS}^{\prime }=\widehat{{\cal A}}_{S}^{\prime }\oplus {\cal V}%
_{rS}^{\prime }  \label{5.6.2}
\end{equation}

If in addition they also satisfy:

4'.- $\widehat{\rho }(\omega )$ is a {\it positive}-valued regular function
of ${\cal S(}{\Bbb R}^{+}{\Bbb )}$, and

5'.- $\rho _{r}(\omega ,\omega ^{\prime })$ is a {\it positive kernel}

\noindent they belong to

\begin{equation}
{\cal A}_{vHS+}^{\prime }=\widehat{{\cal A}}_{S+}^{\prime }\oplus {\cal V}%
_{rS+}^{\prime }  \label{5.6.3}
\end{equation}

Finally, if they verify:

6'.- the normalization condition 
\begin{equation}
\rho ({\Bbb I})=\int\limits_{0}^{+\infty }\widehat{\rho }(\omega {\Bbb )}d%
{\Bbb \omega =}1  \label{5.7}
\end{equation}
then we get a {\bf van Hove state} $\rho \in N\left( {\cal A}_{vHS+}^{\prime
}\right) :$

\begin{equation}
N\left( {\cal A}_{vHS+}^{\prime }\right) =N\left( \widehat{{\cal A}}%
_{S+}^{\prime }\right) \oplus {\cal V}_{rS+}^{\prime }  \label{5.8}
\end{equation}

(There is no normalization in ${\cal V}_{rS+}^{\prime }$ since ${\Bbb I\in }%
\widehat{{\cal A}}_{S+}).$ As before, we will write 
\begin{equation}
\rho =\widehat{\rho }+\rho _{r}  \label{5.8.1}
\end{equation}
and so 
\begin{equation}
\rho \left( a\right) =\widehat{\rho }\left( \widehat{a}\right) +\rho
_{r}\left( a_{r}\right)  \label{5.8.2}
\end{equation}

Let us consider the time evolution: 
\begin{equation}
\left( {\cal U}_{t}\rho \right) (a)=\rho \left( {\cal U}_{t}(a)\right) =%
\widehat{\rho }\left( \widehat{a}\right) +\rho _{r}\left( {\cal U}%
_{t}(a_{r})\right)  \label{5.10}
\end{equation}

We can see that the singular part of the states is {\bf invariant} under
time evolution, while the regular one {\bf fluctuates}. Now, we can compute 
\begin{eqnarray}
\rho _{r}\left( {\cal U}_{t}(a_{r})\right) &=&\int \int \left( {\cal U}%
_{t}(a_{r})\right) (\omega ^{\prime },\omega )\rho _{r}(\omega ,\omega
^{\prime })d\omega d\omega ^{\prime }  \nonumber \\
&=&\int \int e^{i\omega ^{\prime }t}a_{r}(\omega ^{\prime },\omega
)e^{-i\omega t}\rho _{r}(\omega ,\omega ^{\prime })d\omega d\omega ^{\prime }
\nonumber \\
&=&\int \int e^{i(\omega ^{\prime }-\omega )t}\rho _{r}(\omega ,\omega
^{\prime })a_{r}(\omega ^{\prime },\omega )d\omega d\omega ^{\prime }
\label{5.11}
\end{eqnarray}
where the region of integration is the cartesian square of an interval of
the real numbers, according to the spectral hypotesis we have made.

All the functions in the integrand of the r. h. s. of eq. (\ref{5.11}) have
being endowed with the properties listed in the Appendix of paper \cite{L}
(needed in order to satisfy -by large- the hypotesis of the Riemann-Lebesgue
theorem, in the two variables case), and therefore we can conclude: 
\begin{equation}
\lim_{t\rightarrow \infty }\rho _{r}\left( {\cal U}_{t}(a_{r})\right) =0
\label{5.12}
\end{equation}

So, we have proved that: \cite{Rudin} 
\begin{equation}
w^{*}-\lim_{t\rightarrow \infty }\rho _{r}=0  \label{5.13}
\end{equation}
or equivalently: in this simple case {\it any van Hove state becomes
weakly-star diagonal for} $t\rightarrow +\infty .$ This means decoherence in
a purely algebraic framework.

\section{The nuclear *-algebra ${\cal L}\left( {\cal S}_{\Lambda }\left( 
{\Bbb R}^{N+1}\right) \right) $}

Having established (for the simplest case, and when there is some fixed
Hilbert space ${\cal H}$) the decoherence in the energy $H$ we can consider
its generalization for more complex systems, as well as the decoherence in
other commuting constant of the motion $O_{1},...,O_{N}$ as in paper \cite{C}%
, section II.B. Moreover, we would like to show the generalization of the
simplest case of finite discrete indices, considered in that paper, to
finite but otherwise arbitrary kinds of indices.

Now, let the CSCO be $\{H,O_{1},...,O_{N}\}$, where we will suppose -as
usually- that all the observables are bounded or unbounded essentially self
adjoint operators \cite{Reed} of a Hilbert states space ${\cal H}$ having a
common dense domain.

According to the {\bf nuclear spectral theorem} \cite{A. Bohm} \cite{Gelfand}%
, there exists a nuclear space $\Phi $ and a rigging of it with ${\cal H}$ : 
\begin{equation}
\Phi \subset {\cal H}\subset \Phi ^{\times }  \label{4.1}
\end{equation}
such that:

i) all the observables of the CSCO have $\Phi $ as a common dense domain,
and are elements of the algebra ${\cal L}\left( \Phi \right) $ of continuous
linear operators form $\Phi $ to $\Phi .$

ii) there is a $\Phi $-complete spectral resolution of the CSCO, in the
sense that there is a basis of generalized eigenvectors $\left\{ \left|
\omega ,\,o_{1},...,\,o_{N}\right\rangle \right\} \subset \Phi ^{\times }$ ,
and a numerical measure $\mu $ on the spectrum 
\[
\Lambda =\Lambda _{1}\times ...\times \Lambda _{N+1}\subset {\Bbb R}^{N+1} 
\]
of the clausure $\overline{H},\overline{O_{1}},...,\overline{O_{N}}$ of the
elements of the CSCO, such that 
\begin{equation}
\forall \varphi ,\,\psi \in \Phi :\left\langle \varphi ,\psi \right\rangle
=\int\limits_{\Lambda }\left\langle \varphi \mid \omega
,\,o_{1},...,\,o_{N}\right\rangle \left\langle \omega
,\,o_{1},...,\,o_{N}\mid \psi \right\rangle d\mu  \label{4.2}
\end{equation}
where in l. h. s. is indicated the scalar product of $\varphi $ and $\,\psi $
in ${\cal H}$ (with the antilinear factor in the left), whereas in the r. h.
s. $\left\langle \varphi \mid \omega ,\,o_{1},...,\,o_{N}\right\rangle $
means the (antilinear) generalized right eigenvector evaluated in $\varphi $
and $\left\langle \omega ,\,o_{1},...,\,o_{N}\mid \psi \right\rangle $ the
(linear) generalized left eigenvector evaluated in $\psi $. If it happens
that some of the $O_{j}$ have purely point spectra, then the factors of
measure $\mu $ on the $\Lambda _{j}$ will be atomic and the corresponding
integrals in eq. (\ref{4.2}) are really sums. So, this is a generalization
of the pure point spectrum case considered in \cite{C}.

Denoting

\begin{eqnarray}
\Omega &=&\left( \omega ,\,o_{1},...,\,o_{N}\right) \in \Lambda  \label{4.3}
\\
\Omega ^{\prime } &=&\left( \omega ^{\prime },\,o_{1}^{\prime
},...,\,o_{N}^{\prime }\right) \in \Lambda  \label{4.4}
\end{eqnarray}
it is possible to simplify the forthcoming notation. For example, eq. (\ref
{4.2}), now is: 
\begin{equation}
\forall \varphi ,\,\psi \in \Phi :\left\langle \varphi ,\psi \right\rangle
=\int\limits_{\Lambda }\left\langle \varphi \mid \Omega \right\rangle
\left\langle \Omega \mid \psi \right\rangle d\mu  \label{4.2.1}
\end{equation}

Let us consider the set ${\cal S}_{\Lambda }\left( {\Bbb R}^{N+1}\right) $
formed by the restrictions to $\Lambda $ (the spectrum of the CSCO) of all
the functions belonging to the Schwarz space ${\cal S}\left( {\Bbb R}%
^{N+1}\right) $. Being the image of the natural ``onto'' linear map $\Pi $ 
\[
f\longmapsto \Pi (f)=f\mid _{\Lambda } 
\]
${\cal S}_{\Lambda }\left( {\Bbb R}^{N+1}\right) $ can be considered as a
quotient space 
\[
{\cal S}_{\Lambda }\left( {\Bbb R}^{N+1}\right) \cong {\cal S}\left( {\Bbb R}%
^{N+1}\right) /\;Ker\,(\Pi ) 
\]
of the nuclear space ${\cal S}\left( {\Bbb R}^{N+1}\right) $ modulo the
closed linear subspace $Ker\,(\Pi )$. By a wellknown result (\cite{Treves},
Proposition 50.1) this space is nuclear. There are good physical reasons for
such a choice. In fact, any physical observable of a quatum system like
those considered here, decreases fast or even vanishes at infinity, and it
only matters within the limits of its own spectrum, where all the results of
its experimental measures lay.

Now, any observable or ``generalized matrix'' of our system must be a kernel 
$O(\Omega ,\,\Omega ^{\prime })$, and the product must be a generalization
of the product of matrices, such as the composition of linear mappings into
infinite-dimensional vector spaces. Taking these ideas in mind, as well as
the experience left by our previous example, let us consider as the
characteristic algebra 
\begin{equation}
{\cal A}={\cal L}\left( {\cal S}_{\Lambda }\left( {\Bbb R}^{N+1}\right)
\right) \cong {\cal S}_{\Lambda }^{\prime }\left( {\Bbb R}^{N+1}\right) 
\widehat{\otimes }{\cal S}_{\Lambda }\left( {\Bbb R}^{N+1}\right)
\label{4.5}
\end{equation}

By the same arguments as before, ${\cal A}$ is a nuclear *-algebra.
Moreover, it is dual-nuclear, i. e., its dual ${\cal A}^{\prime }$ is also
nuclear. In fact 
\begin{equation}
{\cal A}^{\prime }\cong {\cal S}_{\Lambda }\left( {\Bbb R}^{N+1}\right) 
\widehat{\otimes }{\cal S}_{\Lambda }^{\prime }\left( {\Bbb R}^{N+1}\right)
\label{4.5.1}
\end{equation}

As it was before, let $\widehat{{\cal A}}$ be the abelian subalgebra
generated by this CSCO, and define: 
\begin{eqnarray}
{\cal A}_{vH} &=&\widehat{{\cal A}}\oplus {\cal V}_{r}  \nonumber \\
&=&\left\{ a\in {\cal A\;}/\;K_{a}(\Omega ,\Omega ^{\prime })=\widehat{a}%
(\Omega )\delta (\omega -\omega ^{\prime })+a_{r}(\Omega ,\Omega ^{\prime
})\right\}  \label{4.6}
\end{eqnarray}
where $\widehat{a}(\Omega )$, $a_{r}(\Omega ,\Omega ^{\prime })$ are
complex-valued regular{\it \ functions}. If we only consider symmetric
operators, we add the subindex $``S"$, getting 
\begin{equation}
{\cal A}_{vHS}=\widehat{{\cal A}}_{S}\oplus {\cal V}_{rS}  \label{4.7}
\end{equation}

With regard to functionals, we demand

\begin{eqnarray}
{\cal A}_{vHS+}^{\prime } &=&\widehat{{\cal A}}_{S+}^{\prime }\oplus {\cal V}%
_{rS+}^{\prime }  \nonumber \\
&=&\left\{ \rho \in {\cal A}^{\prime }{\cal \;}/\;K_{\rho }(\Omega ,\Omega
^{\prime })=\widehat{\rho }(\Omega )\delta (\omega -\omega ^{\prime })+\rho
_{r}(\Omega ,\Omega ^{\prime })\right\}  \label{4.8}
\end{eqnarray}
where $\widehat{\rho }(\Omega )$ is a {\it positive}-valued{\it \ }regular 
{\it function}, and $\rho _{r}(\Omega ,\Omega ^{\prime })$ is now a {\it %
positive kernel} verifying: 
\[
\rho _{r}(\Omega ,\Omega ^{\prime })a_{r}(\Omega ^{\prime },\Omega )\in 
{\cal S}\left( {\Bbb R}^{N+1}\times {\Bbb R}^{N+1}\right) 
\]
for any $a_{r}\in {\cal V}_{r}${\it .}

As before, adding the normalization condition 
\begin{equation}
\rho ({\Bbb I})=\int_{\Lambda }\widehat{\rho }(\Omega {\Bbb )}d\mu {\Bbb =}1
\label{4.9}
\end{equation}
we get the {\bf van Hove states}

\begin{equation}
N\left( {\cal A}_{vHS+}^{\prime }\right) =N\left( \widehat{{\cal A}}%
_{S+}^{\prime }\right) \oplus {\cal V}_{rS+}^{\prime }  \label{4.10}
\end{equation}

If $a\in {\cal A}_{vH}$ we have: 
\begin{equation}
\rho (a)=\widehat{\rho }(\widehat{a})+\rho _{r}(a_{r})  \label{4.11}
\end{equation}
and for the $2$-variables case of the Riemann-Lebesgue theorem:

\begin{eqnarray}
\lim_{t\rightarrow \infty }\rho _{r}\left( {\cal U}_{t}(a_{r})\right)
&=&\lim_{t\rightarrow \infty }\int \int e^{i(\omega ^{\prime }-\omega
)t}\rho _{r}(\Omega ,\Omega _{^{\prime }})a_{r}(\Omega _{^{\prime }},\Omega
)d\omega d\omega ^{\prime }  \label{4.12} \\
&=&0  \label{4.13}
\end{eqnarray}
and so 
\begin{equation}
\lim_{t\rightarrow \infty }\rho \left( {\cal U}_{t}(a)\right) =\widehat{\rho 
}(\widehat{a})  \label{4.13.1}
\end{equation}

Thus, we have reached the {\it time-independent} component $\widehat{\rho }(%
\widehat{a})$ of $\rho (a),$ defined by the initial conditions. Therefore,
it would be impossible that another decoherence process would {\it follow}
in order to eliminate the off-diagonal terms of the remaining $N$ dynamical
variables that are {\it still present} in $\widehat{\rho }(\widehat{a})$.
This is because $\omega =\omega ^{\prime }$ in eq. (\ref{4.13.1}), and there
is no Riemann-Lebesgue ``destructive interference'' term as in the integrand
of eq. (\ref{4.12}) for the remaining $N$ variables. Nevertheless, by the
generalized GNS-representation theorem for $\widehat{{\cal A}}$ \cite{Iguri} 
\cite{Canadiense} , if ${\cal H}_{\widehat{\rho }}$ is the
``Hilbert-Liouville'' space obtained by completion of $\widehat{{\cal A}}$
with the pre-hilbertian scalar product ``$(-\left| -\right) "$ defined by $%
\widehat{\rho },$ i. e. 
\begin{equation}
\forall \widehat{a},\widehat{\,b}\in \widehat{{\cal A}}:(\widehat{a},%
\widehat{\,b}):=\widehat{\rho }(\widehat{a}^{*}\widehat{\,b})  \label{4.14}
\end{equation}
there is a {\it pointer representation} : 
\[
\pi _{\widehat{\rho }}:\widehat{{\cal A}}\rightarrow {\cal L}\left( {\cal H}%
_{\widehat{\rho }}\right) 
\]
namely, left multiplication by $\widehat{a}$ 
\begin{equation}
\pi _{\widehat{\rho }}\left( \widehat{a}\right) \left[ \widehat{b}\right]
:=\left[ \widehat{a}\widehat{b}\right]  \label{4.15}
\end{equation}
where $\left[ \widehat{b}\right] $ means the equivalence class correspondent
to $\widehat{b}$ in the completion of $\widehat{{\cal A}}$, such that, for
some {\it normal state ``vector'' (}in our quantum theory it {\it really}
has the status of an operator, like a density matrix{\it ) }of the
representation $\left| \widehat{R}\right) $, belonging to a dense subset $D_{%
\widehat{\rho }}$ of ${\cal H}_{\widehat{\rho }}$ containing $\widehat{{\cal %
A}}$, and contained in the domains of all the essentially self adjoint
``super-operators'' (they act on operators) $\pi _{\widehat{\rho }}\left( 
\widehat{{\cal A}}_{S}\right) $, we have 
\begin{equation}
\forall \widehat{a}\in \widehat{{\cal A}}:\left( \widehat{R}\right| \pi _{%
\widehat{\rho }}(\widehat{a})\left| \widehat{R}\right) =\widehat{\rho }(%
\widehat{a})  \label{4.16}
\end{equation}

So, we are again under the hypotesis of the nuclear spectral theorem for the 
{\bf final pointer CSCO} 
\begin{equation}
\left\{ \pi _{\widehat{\rho }}(H),\pi _{\widehat{\rho }}(O_{1}),...,\pi _{%
\widehat{\rho }}(O_{N})\right\}  \label{4.17}
\end{equation}
of ${\cal H}_{\widehat{\rho }}$ and therefore there exist a nuclear space $%
\Psi $ and a rigging of it with ${\cal H}_{\widehat{\rho }}$ 
\begin{equation}
\Psi \subset {\cal H}_{\widehat{\rho }}\subset \Psi ^{\times }
\end{equation}
such that:

i') all the observables of the CSCO have $\Psi $ as a common dense domain,
and are elements of the algebra ${\cal L}\left( \Psi \right) $ of continuous
linear operators form $\Psi $ to $\Psi .$

ii') there is a $\Psi $-complete spectral resolution of the CSCO, in the
sense that there is a basis of generalized eigenvectors 
\begin{equation}
\left\{ \left| \Theta \right) =\left| \theta _{1},\stackrel{N+1}{...},\theta
_{N+1}\right) \;/\;\left( \,\theta _{1},\stackrel{N+1}{...},\,\theta
_{N+1}\right) \in \Sigma \right\} \subset \Psi ^{\times }  \label{4.18}
\end{equation}
where

\[
\Sigma =\Sigma _{1}\times ...\times \Sigma _{N+1}\subset {\Bbb R}^{N+1} 
\]
is the spectrum of the clausures $\overline{\pi _{\widehat{\rho }}(H)},%
\overline{\pi _{\widehat{\rho }}(O_{1})},...,\overline{\pi _{\widehat{\rho }%
}(O_{N})\text{, }}$and a numerical positive measure $\sigma $ on $\Sigma $,
such that 
\begin{equation}
\forall \widehat{a},\widehat{b}\in \widehat{{\cal A}}\cap \Psi :\left( 
\widehat{a},\widehat{b}\right) =\int\limits_{\Sigma }(\widehat{a}\mid \Theta
)(\Theta \mid \widehat{b})d\sigma =\widehat{\rho }(\widehat{a}^{*}\widehat{b}%
)  \label{4.19}
\end{equation}

In particular, for $\widehat{a}={\Bbb I}$ and $\widehat{b}=\widehat{a}$%
\begin{eqnarray}
\widehat{\rho }(\widehat{a}) &=&\int\limits_{\Sigma }({\Bbb I}\mid \Theta
)(\Theta \mid \widehat{a})d\sigma  \nonumber \\
&=&\left\{ \int\limits_{\Sigma }\widehat{\rho }\left( \Theta \right) \left(
\Theta \right| d\sigma \right\} \left| \widehat{a}\right)  \label{4.20}
\end{eqnarray}
where $\widehat{\rho }\left( \Theta \right) :=({\Bbb I}\mid \Theta )$.

Thus, star-weakly (i. e. when evaluating in $\widehat{a}\in \widehat{{\cal A}%
}$), we have: 
\begin{equation}
\widehat{\rho }=\int\limits_{\Sigma }\widehat{\rho }\left( \Theta \right)
\left( \Theta \right| d\sigma  \label{4.21}
\end{equation}

Eq. (\ref{4.21}) shows that $\widehat{\rho }$ is star-weakly diagonal, but
this time {\it in all its indices.}

Then 
\begin{equation}
\rho =\widehat{\rho }+\rho _{r}=\int\limits_{\Sigma }\widehat{\rho }\left(
\Theta \right) \left( \Theta \right| d\sigma +\rho _{r}  \label{4.22}
\end{equation}
and 
\begin{equation}
w^{*}-\lim_{t\rightarrow \infty }\rho =\widehat{\rho }  \label{4.23}
\end{equation}

The state-dependent basis (\ref{4.18}) in which {\it all} the off-diagonal
components of $\rho $ star-weakly go to zero for $t\rightarrow \infty $, is
called the {\bf final pointer basis}{\it . }In it we have complete
decoherence.

Summarizing:

i.- Decoherence in the energy is produced by the time evolution when $%
t\rightarrow +\infty .$

ii.- Decoherence in the other dynamical variables appears if we choose an
adequate generalized eigen-basis, namely the final pointer basis.

\section{Conclusions}

At the introduction we said that the cause of decoherence is a combination
of certain dynamical properties of the system itself, together with an
unavoidable restriction of the accesible information.

Then, two final comments are in order.

1.- It is necessary that the system has an absolutely continuous evolution
spectrum, which implies a certain degree of complexity (in fact, classically
mixing dynamical systems have this kind of spectra \cite{Reed}). Moreover,
from the use of the Riemann-Lebesgue theorem we may have the feeling that 
{\it all} systems do decohere. While this is theoretically true for $%
t\rightarrow \infty $, it is not so for finite time. This problem is
discussed in paper \cite{C}, and it turns out that systems with an infinite
characteristic time do not decohere, if considered on physical grounds.
Therefore, continuous spectrum and finite decoherence characteristic time
are the dynamical properties needed for the system to decohere.

2.- With respect to decoherence, what Physics finally has to deal with is
not the whole characteristic algebra of the system, but only the {\it actual}
set of its measurable {\it observables}. This is the unavoidable restriction
of the accesible information. Therefore, if we choose to work with an
algebraic formalism, our task would be to find some ``method of
restriction'' of the characteristic algebra, or of its observables, in order
to explain the underlying mechanism behind this kind of phenomena. Ideally,
we would like this restriction to be not too strong as to lead us too far
away of our original system model. On the contrary, we would like to be as
close as possible to the algebra ${\cal A}$. That is why we have chosen an $%
{\cal A}_{vH}$ which is dense in ${\cal A}$.

\end{document}